\documentstyle[aps]{revtex}
\begin{document}
\title { Surface versus crystal-termination effects \\ in the optical properties of surfaces}

\author{ Rodolfo Del Sole and Giovanni Onida }

\address{   Istituto Nazionale per la Fisica della Materia,
 Dipartimento di Fisica dell' Universit\`a di Roma Tor Vergata,
   Via della Ricerca Scientifica, I--00133 Roma,
   Italy}

\par\noindent

\date{\today}

\maketitle
\begin{abstract}

We prove, by realistic microscopic calculations within the sp$^3$s$^*$ Tight Binding method
for GaAs (110) and (100), that the surface optical properties
 are not influenced by long--range
crystal termination effects, and hence that they can be consistently studied considering slabs of 
limited thickness ( 20 -- 30 $\AA$).
The origin of derivative-like and bulk-like lineshapes in Reflection Anisotropy Spectra is also
discussed, analyzing the effects arising from
possible surface-induced reduction, broadening, and shifting of the bulk
spectrum near the surface.

\end{abstract}
\pacs{ PACS numbers: }
\narrowtext
\section{Introduction}

Reflectance Anisotropy Spectroscopy (RAS) and Surface 
Differential Reflectance (SDR) are surface-sensitive optical 
techniques, and are used to obtain information on the atomic 
and electronic structures of surfaces \cite{halevi}. In the early times of these 
spectroscopies, the spectra were explained in terms of 
transitions across surface states, although this view has been 
contrasted by calculations, showing that surface-geometry effects 
could also determine the 
spectra through surface perturbations on the optical matrix 
elements of transitions across
bulk states \cite{selloni}. Now the attitude seems to be reversed: after 
realizing 
that many RA lineshapes are similar to the imaginary part of the 
bulk dielectric function, or to its energy derivative, it is growing 
the belief that surface optical spectra are mostly determined by 
bulk effects, and therefore not very useful as a tool of surface 
characterization. In this paper, we discuss the 
origin of these bulk-like features, and at the same time emphasize 
the presence in optical spectra of other features, more related to 
the surface structure.

In 1996, Rossow, Mantese and Aspnes \cite{rossow} recognized that RAS 
data 
on chemically saturated surfaces generally resemble the energy 
derivative lineshapes of the corresponding bulk spectra, 
$dIm[\varepsilon_b 
(\omega)]/d(\omega)$, while  
surfaces with unsaturated dangling bonds (DBs) often yield RAS 
lineshapes resembling the bulk spectrum, $Im[\varepsilon_b 
(\omega)]$. They explained the latter lineshapes in terms of 
surface-
induced changes of the electron-hole interaction and of local 
fields, while derivative-like spectra were explained in terms of 
surface-perturbations on the energies of bulk states. From these 
findings, 
they inferred the occurrence of shorter lifetimes of electrons and 
holes near 
the surfaces. We show here that this deduction is not necessary. 
Furthermore, we demonstrate the existence of other 
mechanisms able to produce energy-derivative lineshapes.

A further step along the way of attributing most RAS and SDR 
features to bulk effects has been done by Uwai and Kobayashi 
(UK) in 1997 \cite{UK}. They measured surface photoabsorption (SPA) 
spectra for different conditions of the GaAs (001) and (111) 
surfaces, from which the changes of the surface dielectric tensor 
 were extracted. The imaginary parts 
of such changes have peaks at 2.6-3 eV and at 4.5-4.7 eV, 
close to the main structures, $E_1$ and $E_2$,
 of the bulk dielectric function. The lineshapes are 
similar to the imaginary part of the bulk dielectric function in the 
case of the (001) surface, and to its derivative for the (111) surface. 
UK conclude that these two peaks are not due to transitions 
involving surface states, but to modified bulk electronic 
transitions.
They claim that the surface termination effect, first 
discussed by one of the present authors in 1975 \cite{delsole}, is responsible 
for 
the occurrence of bulk-like features in surface spectra. According 
to UK, this effect mostly consists in a reduction of the
polarizability below the surface, arising from the quenching of bulk-state 
wavefunctions near the surface, due to their vanishing
outside the crystal. This might be a long-range 
effect, extending one hundred Angstroms below the surface, which 
might be hardly included in slab calculations. We show here that, 
although the crystal-termination effect is in fact present, the way 
it has been described by UK is rather naive; not a bare reduction 
of the polarizability, but a distortion of its lineshape must occur 
(and indeed occurs), to produce a nonvanishing RAS or SDR signal. 
However, the resulting effect is by no means of long range, and is 
in fact included in slab calculations. Moreover, while the crystal-
termination effect often yields derivative-like lineshapes, we have 
not found bulk-like spectra arising from it.

The GaAs(110) surface is a good test case for our calculations and 
discussions, because of its well defined atomic structure and for 
the occurrence of (modest) surface effects partially overlapping in 
energy with (predominant) bulk effects \cite{manghi,pulci}. The As-rich 
GaAs(100) $\beta 2$(2x4) surface will also be considered. 

\section{Theory}
We calculate the surface contribution to reflectance, that is its 
relative deviation with respect to Fresnel formulas, according to 
the three-layer model \cite{aspnesmcint}:

\begin{equation}\label{rs}
\frac{\Delta R_{i}}{R} = \frac{4 \omega}{c} \cos \theta d
\ Im \left( \frac{ 
\varepsilon_{si}(\omega) - 
\varepsilon_b(\omega)}{\varepsilon_b(\omega)-1}
\right)  ,
\end{equation}
\noindent
for s-light polarized parallel to the i-direction (i = x or y) in the 
surface plane, 
where $\theta$ is the angle of incidence, d the depth of the 
surface layer, $\varepsilon_{si}(\omega)$ is the ii diagonal 
component of the surface-layer dielectric tensor, and 
$\varepsilon_{b}(\omega)$ is the isotropic bulk dielectric function. 
For p--light incident in the iz--plane, the anisotropic three--layer model yields
\cite{halevi}: 

\begin{equation}\label{rp}
\frac{\Delta R_{i}}{R} = \frac{4 \omega}{c} \cos \theta d
\ Im \left( \frac{ 
(\varepsilon_{si}(\omega) - 
\varepsilon_b(\omega))(\varepsilon_b(\omega)-sin^2 
\theta)+\varepsilon^2_b(\omega) sin^2 \theta 
(1/\varepsilon_{sz}(\omega)-
1/\varepsilon_b(\omega))}{(\varepsilon_b(\omega)-
1)(\varepsilon_b(\omega) cos^2 \theta - sin^2 \theta)}
\right)  .
\end{equation}
\noindent
The surface-layer dielectric tensor, 
assumed to be diagonal, is obtained by subtracting 
 the bulk dielectric function
from the 
calculated 
slab dielectric tensor, with a suitable choice of 
the surface-layer depth, d. The reflectivity for s light comes out to 
be independent of the choice of d, coincident with the 
microscopic formulas not relying on the three-layer model. This model is 
instead needed to obtain p-light reflectivity by avoiding the 
computationally very demanding inversion of the 
dielectric susceptibility tensor.

\section{Results}
We start by calculating the normal-incidence reflectance 
anisotropy (RA), $2(R_x - R_y)/(R_x + R_y)$, 
of GaAs(110). The latter is the cleavage surface of GaAs, and,
despite being not reconstructed, undergoes large relaxations.
Its equilibrium structure, known as the ``rotation-
relaxation model'', is well known both from the experimental and the 
theoretical sides: the surface As atoms relax
toward the vacuum, and Ga atoms move in the opposite direction,
recovering a quasi-planar sp$^2$ 
bonding with their three As neighbors \cite{rotationrelax}.
We represent the surface using a slab of 31 atomic layers, where
the actual atomic positions are taken from a Car-Parrinello 
total energy minimization \cite{cpgaas110}.
Since the slab has two equivalent surfaces, 
the computed slab polarizability must be divided by two.
We then also consider a polar surface of  GaAs: the As-rich 
(100)  $\beta 2$(2x4).  The latter is 
known to be the stable reconstruction for this surface \cite{struct2x4},
and is a regular array of two As dimers and two dimer vacancies 
(the unit cell contains only two As dimers), 
aligned along the [$\bar{1}10$] direction. Also in this case 
the actual atomic positions are taken from a Car-Parrinello
total energy minimization \cite{gero}.
In the case of GaAs(100), 
since geometry does not allow to build a slab with two
equivalent surfaces, the calculation is done for a system with only
one surface reconstructed, i.e. by including a real--space cutoff function
(a squared cosine, approaching one on the interesting surface and zero on the 
other), 
in the optical transition probability calculations,
to eliminate the contribution of the back surface\cite{nota1}.
% (in fact,
% the two surface contributions to the diagonal matrix 
% elements of $\varepsilon_{i,j}$
% always sum each other: cancellations can only occur for the 
% non--diagonal elements, which are not considered here).
The electronic states of the slab, as well as of 
those of bulk GaAs, are calculated according to the $sp^3s^*$ tight-
binding method, as in Ref. \cite{selloni}. The As--As 
tight--binding interaction parameters are those of ref. \cite{reilly}.
The imaginary part of the slab 
dielectric function is obtained by considering transitions at a  number of k 
points in the irreducible part of the two-
dimensional 
Brillouin Zone (IBZ). The first issue we address is the number of k 
points which are needed to obtain a good convergence. In Fig. 1 
we show the RA of GaAs(110) calculated with 256, 1024 and 
4096 special k-points in the IBZ; the curves corresponding to the 
first two cases are clearly distinct from each other. The 
calculation with 4096 k points, instead, is almost coincident with 
that with 1024 k points. This means that 1024 k points are 
needed to achieve full quantitative convergence of the GaAs(110) RA. 
This result might be a peculiar property of this and similar 
surfaces. In the case of the GaAs(100) $\beta 2$(2x4) surface,
a good convergence of the spectrum is already obtained 
using a  number of k-points equivalent to 
64 in the (1x1) surface cell. 
Similar calculations carried out on Si(110):H show that the RA spectrum is already 
converged with 64 k points (usually, calculations are made with 64 k points 
or less \cite{selloni,manghi,pulci}, since,
even with 
a well converged k-point summation,
only qualitative accuracy can be 
achieved, due to the neglect of excitonic and local-field effects.)

The calculated RAS for GaAs(110) is qualitatively similar to previous 
calculations, carried out using tight-binding or ab-initio methods, 
and to 
experiments \cite{manghi,pulci,esser}. The peak at about 2.9 eV 
embodies a substantial
contribution of transitions across surface states or resonances (at variance 
with Ref. \cite{manghi}, but in agreement with references \cite{pulci} and 
\cite{esser}), 
while the higher-energy structures are essentially due to 
transitions across surface-perturbed bulk states. The main effect 
of the k-point convergence achieved in the present calculation 
was to reduce the intensity of the dip just above the 2.9 eV peak 
and of the subsequent structures.

Having achieved quantitative convergence with respect to the 
number of k points, we can look now at the convergence with 
respect to the number of layers, which is the main interest here, 
since changes in lineshapes occurring for very thick slabs would 
indicate the presence of the long-range effect assumed by UK \cite{UK}. In 
Fig. 2a we show the GaAS(110) RA calculated using slabs of 11, 31 and 93 
layers and 1024 k points.
 The latter two curves are almost indistinguishable, while 
the 11-layer curve is also close to them. This means that the 
calculation has already converged with 31 layers, and that the 
aforementioned long-range effect does not occur.
The same is true for the polar (100) surface: in fig 2c we show
the calculated RA for the GaAs(100) $\beta 2$(2x4) surface,
where small differences show up using slabs of 16, 20 and 40 layers.
The slow convergence with slab thickness 
observed in calculations \cite{pulci} when a 
smaller set of k points is used is therefore due to the error 
caused by the small number of k points, which randomly varies
with the number of planes.
 This is nicely demonstrated by Fig. 2b,
where the same series of slabs as in Fig 2a 
(11--31--93 layers) has been used 
to compute the GaAs(110) RAS spectrum
with a set of 256 k--points.

In view of the
different structures of Eqs. (1), for s light, and (2), for p light, one
could speculate that the long-range effect might cancel in the
former case, and appear in the latter. To check this possibility, we
present in Fig. 3 the surface contribution to p-light reflectance
calculated using the anisotropic three-layer model at an angle of
incidence of 60 degrees. Again no difference is present between
the curves calculated with 31 and 93 layers, definitely showing
that surface optical properties are well converged
with slabs of 31 layers.
The present results 
increase our confidence in slab calculation, not only since very 
long-range effects, which can be hardly embodied therein, are 
excluded, but also because thin slabs, as the 11 and 16-layers ones,
which 
are the only ones that can be afforded in ab-initio calculations \cite{pulci}, 
already yield rather good results.

Let us discuss now in more detail the crystal-termination effect 
invoked by UK \cite{UK}. It has been firstly addressed by one of the present 
authors in connection with the reflectivity at the direct minimum gap of a 
semiconductor \cite{delsole}. By disregarding the microscopic structure of 
the 
surface, and describing it just as an infinite potential barrier 
preventing electrons from escaping into vacuum (the crystal 
termination), the reflectivity was obtained starting from the 
wavefunctions calculated according to the effective-mass 
approximation. Since the wavefunctions must vanish at the 
crystal-termination plane, the envelope plane-waves occurring in 
an infinite crystal are replaced by sine-type standing waves. 
When looking at the imaginary part of the local dielectric function, 
$Im[\varepsilon(z, \omega)]$, this yields a region below the surface 
where this quantity is smaller than in the bulk crystal. The depth 
of such a region is of the order of $\pi /k_z$,  $k_z$ being the 
largest wavevector of the relevant transitions. In the 
case of the minimum gap,
 hence,  $k_z=[2m^*(\hbar \omega - E_g)/\hbar ^2]^{1/2}$, where $m^*$ 
is the reduced electron-hole effective mass and $E_g$ the direct-
gap energy. By taking an effective mass of 0.1 and $\hbar \omega 
- E_g$ as 0.1 eV, we estimate this depth to be of the order of 60 
Angstrom. This is the quenching of the (bulk) dielectric function 
below the surface, that UK assume as the most important effect. 
However, crystal termination affects the optical properties also in 
another way: since the matrix elements of the momentum 
operator must be calculated between the surface-perturbed 
wavefunctions (sine-type in the case discussed here), they are 
different from those of the infinite crystal. More explicitly, $k_z$ 
is no longer conserved in subsurface optical transitions; the 
breaking of this selection rule yields spectra distorted with 
respect to (namely, broader than) the corresponding bulk spectra, 
acting as 
an additional broadening localized near the surface. We will show 
below that this additional broadening is the most important 
crystal-termination effect influencing the surface optical 
properties of GaAs. 

Differently from the case discussed above, the main structures of 
bulk spectra, which yield the most prominent bulk-related 
structures in RAS and SDR, are due to transitions at saddle-points 
of the joint density of states. The
characteristic $k_z$ 
at saddle points is much larger than at the direct gap, because a 
large region of Brillouin Zone is available for optical transitions at 
the saddle-point energy. 
For instance, vertical transitions along all the $\Lambda$ line are responsible 
for the $E_1$ structure in GaAs.
Hence the largest $k_z$ is of the order of the BZ 
boundary, $\pi/a$, and the depth of the surface-perturbed region is
of the 
order of the lattice constant, $a$. This explains why we have not 
found in 
Figs. 2 and 3 any indication of long-range effects close to saddle-
point 
energies.

The simplest model of 
quenching is to assume that the polarizability is completely 
suppressed within some depth d below the surface. This, however, 
would be equivalent to shift the surface by d, and would not give 
any contribution to the reflectance. Hence we consider a slightly 
different model, where the polarizability is partly quenched, say 50 percent,  
in a depth $d$.
In practice we assume, within the depth $d$, a surface dielectric function
of the form:

\begin{equation}
 \varepsilon_{si}(\omega) = {\rm f}_i \cdot
\varepsilon_{b}(\omega - \Delta \omega_i, \gamma_i) 
\end{equation} 
where $\varepsilon_{b}(\omega)$ is the bulk dielectric function,
${\rm f}_i$ ($\leq 1$) represents the quenching, $\gamma_i$ is the 
broadening (possibly different from the bulk one), and 
$\Delta \omega_i$ is a possible frequency  shift.
When $\Delta \omega_i = 0$, $\gamma_i= \gamma_{bulk}$, and
${\rm f}_i =1 $, $ \varepsilon_{si}(\omega)$ coincides with the 
bulk dielectric function. Taking ${\rm f}_i < 1$ with 
$\gamma_i= \gamma_{bulk}$ and $\Delta \omega_i = 0$ would 
not modify the s-light 
reflectivity, since the numerator and denominator in equation (1) 
are proportional to each other, and hence  the fraction is a real 
number, with vanishing imaginary part. However, this is not the 
case for p-light 
reflectivity, which may undergo some change. The full line in 
Fig. 4a shows the surface contribution calculated in this way. It is 
clear from the figure that this model has no relation with the 
output of the slab calculation (dashed line); hence the pure quenching effect 
cannot account for the  surface contribution to reflectance.

We consider next the pure broadening model, i.e. 
${\rm f}_i =1 $, $\Delta \omega_i = 0$, and
$\gamma_i > \gamma_{bulk}$. Now the surface is 
assumed to have the same dielectric function as the bulk has, but 
with broader lineshapes, as a consequence of the breaking of the 
$k_z$-conservation near the surface. In Fig. 4b
we show the 
surface contribution to the reflectivity of normally incident light,
with polarization perpendicular and parallel to  the $[1\bar 1 0]$ chains, 
as calculated from the 
slab polarizability (dashed and dotted lines), and according to the broadening 
model (full line). We assume a broadening of 100 meV at the surface,
while it is 30 meV in the bulk.
 The curves are rather similar,  
showing lineshapes resembling the energy-derivative of the 
imaginary part of the dielectric function. Also the surface-state 
related peak at about 2.9 eV is embodied in the broadening-model 
spectrum (this occurs only because the peak mentioned above 
overlaps in energy with the $E_1$ bulk structure around 3 eV).
However, the differences between the broadening model
and microscopic calculations, which seem to be small in this spectrum, 
become very large in the RA spectrum, shown in Fig.  4c.
Here the dashed line is obtained as in Fig. 1, that is from the 31-layer slab 
calculation. We can produce a RA-curve according to the 
broadening 
model by assuming that the depth where the dielectric function is 
broader than in bulk is different for the two polarizations, or, 
in an equivalent manner, that also some quenching of the dielectric function 
occurs (${\rm f}_i < 1$), whose amount depends on the direction of light 
polarization (${\rm f}_x \ne {\rm f}_y$). By 
assuming a suitable depth or quenching difference, we produce 
the full line in Fig. 4c,
which is of course proportional to the continuous line in Fig. 4b.
The two curves in Fig. 4c
are markedly different, 
although the peak at 2.9 eV (the only spectral feature
related to transitions across surface states!) is present in both curves.

Hence we can conclude by recognizing the occurrence
 of bulk- derivative-like features
 in surface 
optical spectra calculated for a given polarization of light,
 due to 
the broader lineshape of the dielectric function near the surface. 
This broader lineshape is due to the breaking of the $k_z$ 
conservation rule (namely, it is a crystal-termination effect), is 
included in slab calculations, and does not imply a shorter lifetime 
of electrons near the surface than in bulk. When anisotropy difference 
spectra are taken for GaAs(110), however, these features largely 
cancel, so that 
the surviving RA has about no relation to the broadening model.
Of course, such cancelation may be smaller at other surfaces, so 
that derivative-like lineshapes may be present in RAS and SDR 
spectra.

As a last point, we can assume that the (bulk) dielectric function 
near the surface
can undergo small shifts of peak positions ($\Delta \omega_i \ne 0$), 
in addition to broadening and quenching. To this aim it is not needed, as 
assumed in Ref.\cite{rossow}, that electrons and holes excited in optical 
transitions are kept close to the surface by their short life-times, in 
order to be shifted in energy by the surface potential. The 
required small shifts of the peaks of the surface dielectric function 
may be produced by the surface-perturbation on the 
wavefunctions and, consequently, on the local polarizability.
It is a matter of fact that the layer-projected density-of-states may be 
different from the bulk one. The same, of course, can occur 
for the z-dependent dielectric function \cite{bacheletss}, whose average
over the first few layers yields the surface dielectric function.

We have tried to obtain bulk-like difference (RAS or SDR) spectra 
by suitably varying ${\rm f}_i$, $\gamma_i$,   and
$\Delta \omega_i$, i.e. by
shifting, broadening and quenching the bulk spectrum.
By varying the parameters above,
we often obtained derivative-like spectra, never obtained bulk-
like spectra, and sometimes hybrid spectra (see Fig. 5a, full line).
 It is worth to notice that this hybrid spectrum is rather similar
(although energy shifted) to the
microscopically calculated RAS spectrum, also shown in Fig. 5a (dashed line).
For some choice of the parameters we got difference spectra
{\it approximately} bulk like, that is showing peaks 
close to the two bulk critical-point energies, but, differently from 
the bulk spectrum, with a negative region in between 
(Fig. 5b, full line). 
A similar RA spectrum is the result of a realistic TB slab 
calculation carried out for another GaAs surface, the (polar) (100) $\beta 2$(2x4).
The calculated RAS is shown in 
Fig. 5b, by the short--dashed line, 
while the experimental 
spectrum,  more similar to the bulk one, corresponds to the
long--dashed line\cite{esser98}. 
This suggests that 
the 
surface-exciton and surface local-field effect may be determinant 
to yield bulk-like surface spectra. A recent calculation for 
Si(110):H \cite{mendoza}, where the experimental RA lineshape is bulk like 
\cite{aspnessi110},
 shows indeed that the surface local-field effect, treated 
therein according to the point-dipole approximation, is crucial to 
obtain a bulk-like theoretical lineshape.

\section{Summary and Conclusions}

To summarize, we have shown that surface effects on optical 
properties of GaAs are localized in a few monolayers below the surface. 
As it has been discussed in section III, 
these results do not depend
explicitly on the particular system considered,
suggesting a more general validity, i.e. indicating that for a wide 
class of semiconductor surfaces
the surface effects on optical
properties, including the crystal termination effect, are well 
described using slabs of a few tens of monolayers. 
In the absence of peculiar features due to surface states, the crystal-
termination effect can be phenomenologically modeled as a shift, broadening and 
reduction of the bulk spectrum. Many combinations of these parameters 
yield 
surface spectra resembling the derivative of 
the bulk absorption spectrum. It must be emphasized that the 
amounts of shift, broadening and reduction are ultimately 
determined by the microscopic structure of the surface; 
furthermore, these bulk-derived structures coexist with 
transitions directly involving surface states. After subtraction of 
individual spectra to obtain RAS or SDR spectra, the resulting 
lineshape can be qualitatively different from a derivative-like 
lineshape, as in the case of GaAs(110). On the other hand,
{\it approximately} bulk-like spectra are obtained
for some values of the parameters. However, {\it truly} 
bulk-like spectra can hardly be obtained in terms of the 
crystal--termination effect, and they did not even occur as results of our 
realistic slab calculations. Hence, many-body effects like the 
surface-exciton or the surface-local field effect, not included in the 
one-electron theory, seem to be determinant to obtain truly bulk-like 
lineshapes. 

In conclusion, we agree with Rossow et al. \cite{rossow} and with UK 
\cite{UK} that some features
of surface spectra originate from transitions across bulk states. 
Surface termination effects, however, involve a more complex mechanism
than that described by  UK. In fact, the 
broadening of the bulk dielectric function near the surface is the most
important crystal termination effect, due to the breaking of $k_z$
conservation at surfaces. It does not imply, however, that
photogenerated electrons and holes have shorter lifetimes than
in the bulk. This effect yields derivative-like lineshapes,
as those obtained at many chemically-saturated surfaces.

Finally, we would like to stress that many effects concur to determine
surface optical properties. It is not possible to interpret optical
spectra of all surfaces in terms of a single effect, either the
crystal termination effect, or transitions across surface states. Caution
must also be used in assigning spectral features to bulk-state
transitions uniquely because of their energy positions, as exemplified by the
case of GaAs(110), where we found that the main peak of the calculated spectrum,
occurring almost at the same energy as the $E_1$ bulk feature, contains
a substantial contribution of transitions across surface states.

\section{acknowledgements}

We thank D.E. Aspnes for stimulating our interest in the problem,
and A.I. Shkrebtii
for his contributions to the tight-binding code used in this work. The 
calculations were performed on a parallel platform  (Cray T3D) at 
the Interuniversity Consortium of the Northeastern
Italy for Automatic Computing (CINECA), under the INFM parallel
computing initiative, account {\it cmprmpi0}.
This work has been finantially supported in part by the Italian Ministery of 
the University and Scientific Research (MURST--COFIN 97).

\bigskip

\begin{figure}
\caption{ Calculated Reflection Anisotropy Spectrum 
of the GaAs(110) surface,
and its convergence with respect to the Brillouin zone sampling. Full line:
 4096 k--points in the irreducible wedge of the Surface  Brillouin zone; 
long--dashed line: 1024 k--points; dotted line: 256 k--points. In the latter
 case, full convergence has not yet been reached.}
\label{fig_conv_kpoints}
\end{figure}

\begin{figure}
\caption{  a) Convergence of the theoretical RA spectrum of GaAs (110),
calculated using 1024 k points, with
respect to the thickness of the slab (number of atomic layers) used in the 
calculation. Full line: 93 layers; long--dashed line: 31 layers; short--dashed
 line: 11 layers. The results for 31 and 91 layers are almost identical. 
b) The same, calculated using 256 k points. 
c) Convergence of the theoretical RA spectrum of GaAs (100) $\beta 2$(2x4),
calculated using 8 k points (equivalent to 64 in the (1x1) surface cell), with
respect to the thickness of the slab. Full line: 40 layers; long--dashed line:
 20 layers; short--dashed line: 16 layers. }
\label{fig_conv_layers}
\end{figure}

\begin{figure}
\caption{ Surface contribution to reflectance for p--light incident 
in the y-z plane at 60 degrees, for GaAs(110)  slabs of 31 layers (dashed line)
and 91 layers (full line), using 1024 k-points in the SBZ.
The y direction is parallel to the chains in the surface plane, while z is
perpendicular to the surface.}
\label{fig31-93_60degrees}
\end{figure}

\begin{figure}
\caption{ a) Microscopic calculation (110 slab) of the surface contribution
 to reflectance for p--light incident in the y-z plane
at 60 degrees (dashed line), compared with the results of the  ``quenching
 model'' (full line);
 b) Microscopic (slab) calculation of the surface contribution
 to reflectance for normally incident light polarized along the (110) chains,
 ( dotted line) and perpendicularly to them ( dashed line), computed with
 a 31-layers GaAs(110)
slab and 1024 k--points in the ISBZ, in comparison with the results of the 
'broadening model' (full line); 
 c) ``Broadening model'' results for the RA spectrum of
GaAs (full line), compared with the microscopic calculation of Fig. 1}
\label{fig_quench_broad}
\end{figure}

\begin{figure} 
\caption{ a) full line: GaAs RAS results
 from the quenching--broadening--shifting model, with the following choice of
parameters: $ {\rm f}_x = 1$, $ {\rm f}_y = 0.5$,
$\gamma_x$ = 0.1 eV, $\gamma_y$ = $\gamma_{bulk}$ = 0.03 eV,
  $ \Delta \omega_x = \Delta \omega_y = 0.1 eV $ (see text). 
Dashed line: microscopic calculation for the (110) slab, 
with  31 layers and 1024 k--points.
b) long-dashed line: experimental RAS data for GaAs (100) $\beta 2$(2x4),
 from ref.{\protect\cite{esser98}}. Short-dashed line: microscopic
calculation for the same surface.
Full line: quenching--broadening--shifting model, with the following choice of
parameters: $ {\rm f}_x = {\rm f}_y = 1$, $\gamma_x = \gamma_y = 0.1 eV$,
$\gamma_{bulk} = 0.03 eV$, $\Delta \omega_x = -0.1 eV$, $\Delta \omega_y = 0$.}
\label{fig_mixed}
\end{figure}

\bigskip

\begin{references}

\bibitem{halevi} R. Del Sole, {\it Reflectance spectroscopy - theory}, in {\it 
Photonic probes of surfaces}, edited by P. Halevi, Elsevier, Amsterdam, (1995), p. 131.
\bibitem{selloni} A. Selloni, P. Marsella and R. Del Sole, Phys. Rev. B {\bf 33}, 
8885 (1986)
\bibitem{rossow} U. Rossow, L. Mantese, and D. E. Aspnes, 
 Proc. 23d Int. Conf. on the Physics of Semiconductors, 
 Berlin, World Scientific, page 831 (1996). 
\bibitem{UK} K. Uwai and N. Kobayashi, Phys. Rev. Lett. {\bf 78}, 959 (1997)
\bibitem{delsole} R. Del Sole, J. Phys. C {\bf 8}, 2971 (1975)
\bibitem{manghi} F. Manghi, R. Del Sole, A. Selloni, E. Molinari, Phys. Rev. B 
{\bf 41}, 9935 (1990)
\bibitem{pulci} O. Pulci, G. Onida, R. Del Sole, and A. I. Shkrebtii, Phys. Rev. B {\bf 58}, 1922 (1998)
\bibitem{aspnesmcint} J.D.E. McInthire and D.E. Aspnes, Surf. Sci. {\bf 24}, 417 
(1971)
\bibitem{rotationrelax} A. Kahn, Surf. Sci. Rep. {\bf 3}, 193 (1983).
\bibitem{cpgaas110} see, e.g., R. Di Felice, A. I. Shkrebtii, F. Finocchi, C. M. Bertoni, 
and G. Onida, J. of Electron Spectroscopy and Related Phenomena {\bf 64/65}, 697 (1993)).
\bibitem{struct2x4} See for instance: A.R Avery, C.M. Goringe, D.M. Holmes, J.L. Sudijono, 
and T.S. Jones, Phys. Rev. Lett {\bf 76}, 3344 (1996); J. G. Schmidt and F. Bechstedt, 
Surf. Sci. Lett {\bf 360} L473 (1996), and references therein. 
\bibitem{gero} J. G. Schmidt and F. Bechstedt,
Surf. Sci. Lett {\bf 360} L473 (1996).
\bibitem{nota1} This cutoff function has also been used to demonstrate that
the absence of long--range effects, discussed below, is not due to 
some cancellation between contributions from  the two opposite surfaces of the slab.
\bibitem{reilly} E. P. O'Reilly and J. Robertson, Phys. Rev. B {\bf 34},8684 (1986).
\bibitem{esser} N. Esser, N. Hunger, J. Rumberg, W. Richter, R. Del Sole, A. I. 
Shkrebtii, Surf. Sci. {\bf  307/309}, A 1045 (1994)
\bibitem{bacheletss} M. Altarelli, G. B. Bachelet, V. Bouche', and R. Del Sole, 
Surf. Sci. {\bf 129}, 447 (1983); see, in particular, Fig. 3.
\bibitem{esser98} A. I. Shkrebtii, N. Esser, W. Richter, W. G. Schmidt, F. 
Bechstedt, A. Kley, and R. Del Sole, Phys. Rev. Lett. {\bf 81}, 721 (1998)
\bibitem{mendoza} B. S. Mendoza, R. Del Sole, and A. I. Shkrebtii, Phys. Rev. 
B {\bf 57}, R12709 (1998)
\bibitem{aspnessi110} T. Yasuda et al., J. Vac. Sci. Technol. A {\bf 12}, 1152 
(1994)
\bibitem{nota2} Of course, long--range effects are indeed present when macroscopic
electric fields build up due to the charging of the surface. This 
occurs, for instance, in the space charge region of semiconductors.
The resulting effects on the reflectance, however, is well 
understood in terms of the Frantz--Keldysh effect (see, for instance,
Victor Rehn, Surf. Sci. {\bf 37}, 443 (1973);
J. W. Grover and P. Handler, Phys. Rev. B {\bf 9}, 2600 (1974),
and references therein
\end{references}
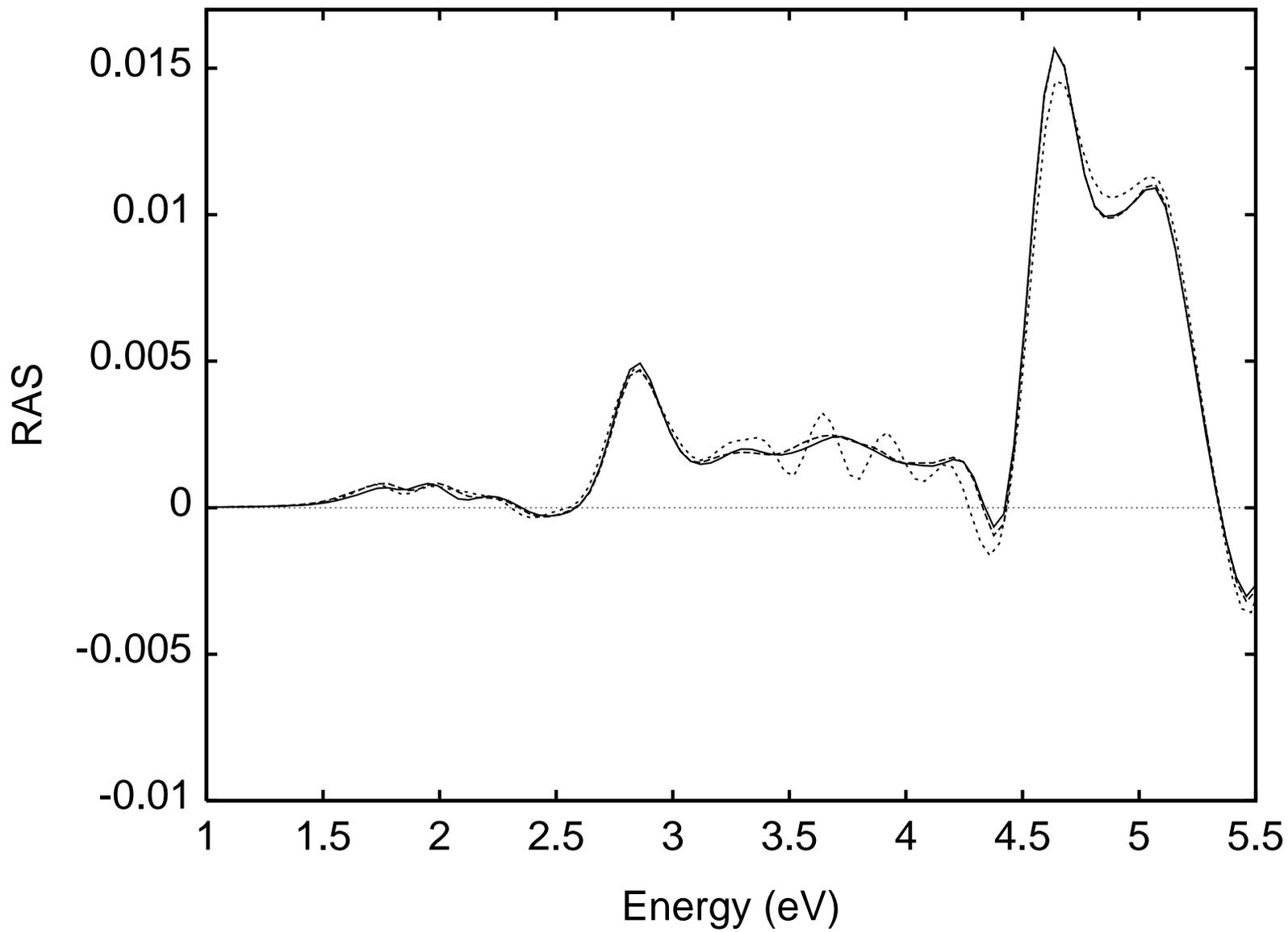
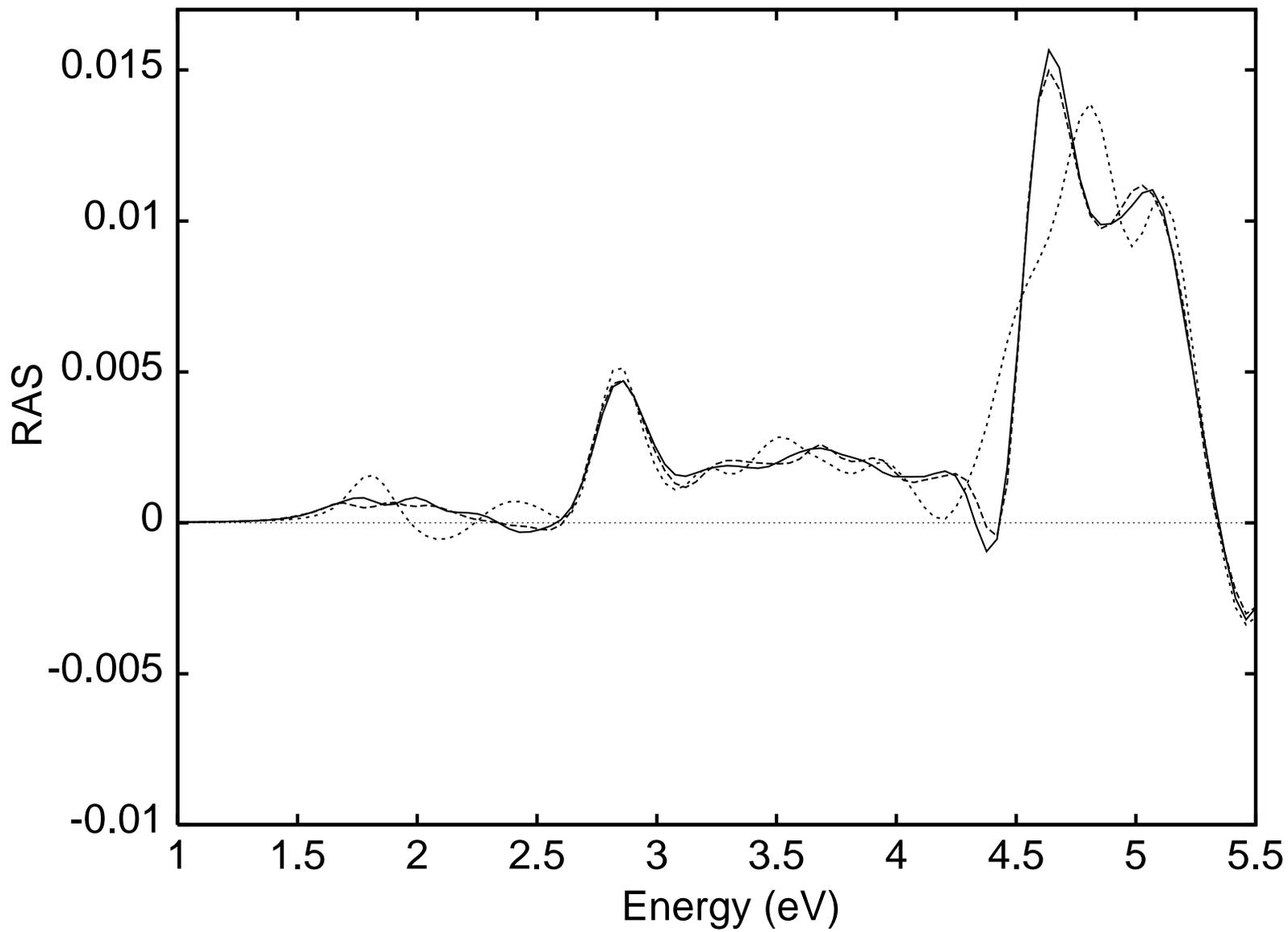
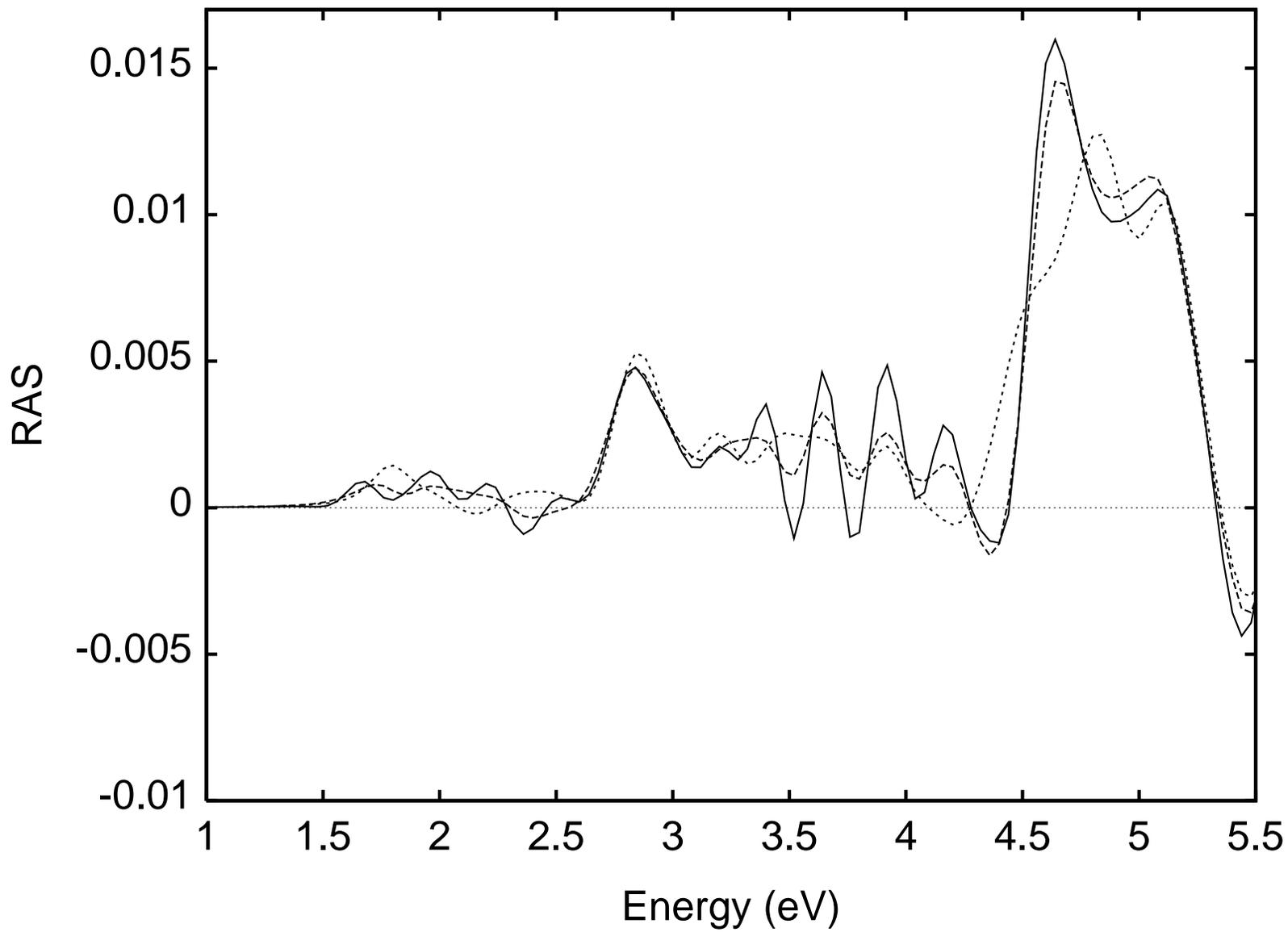
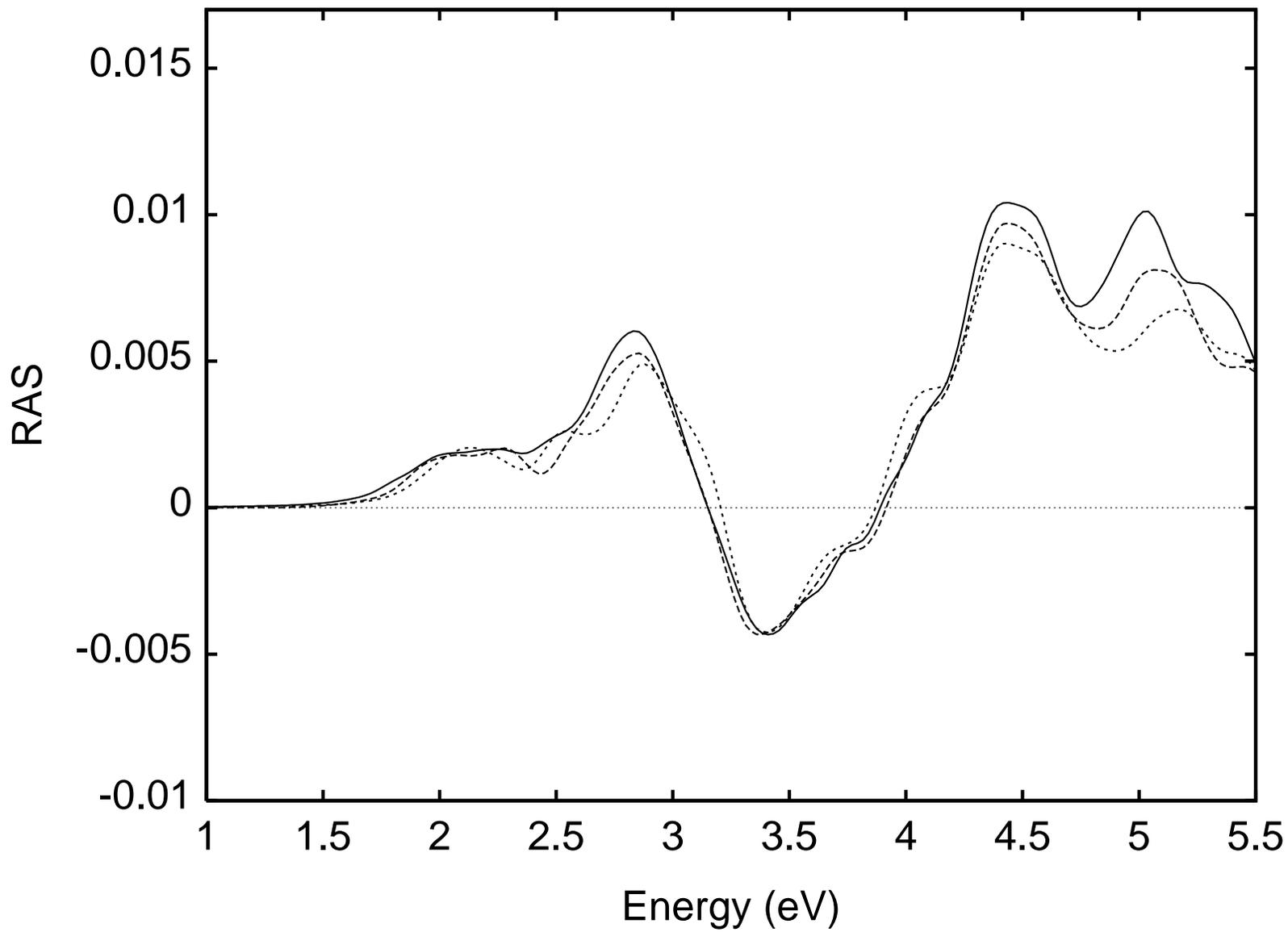
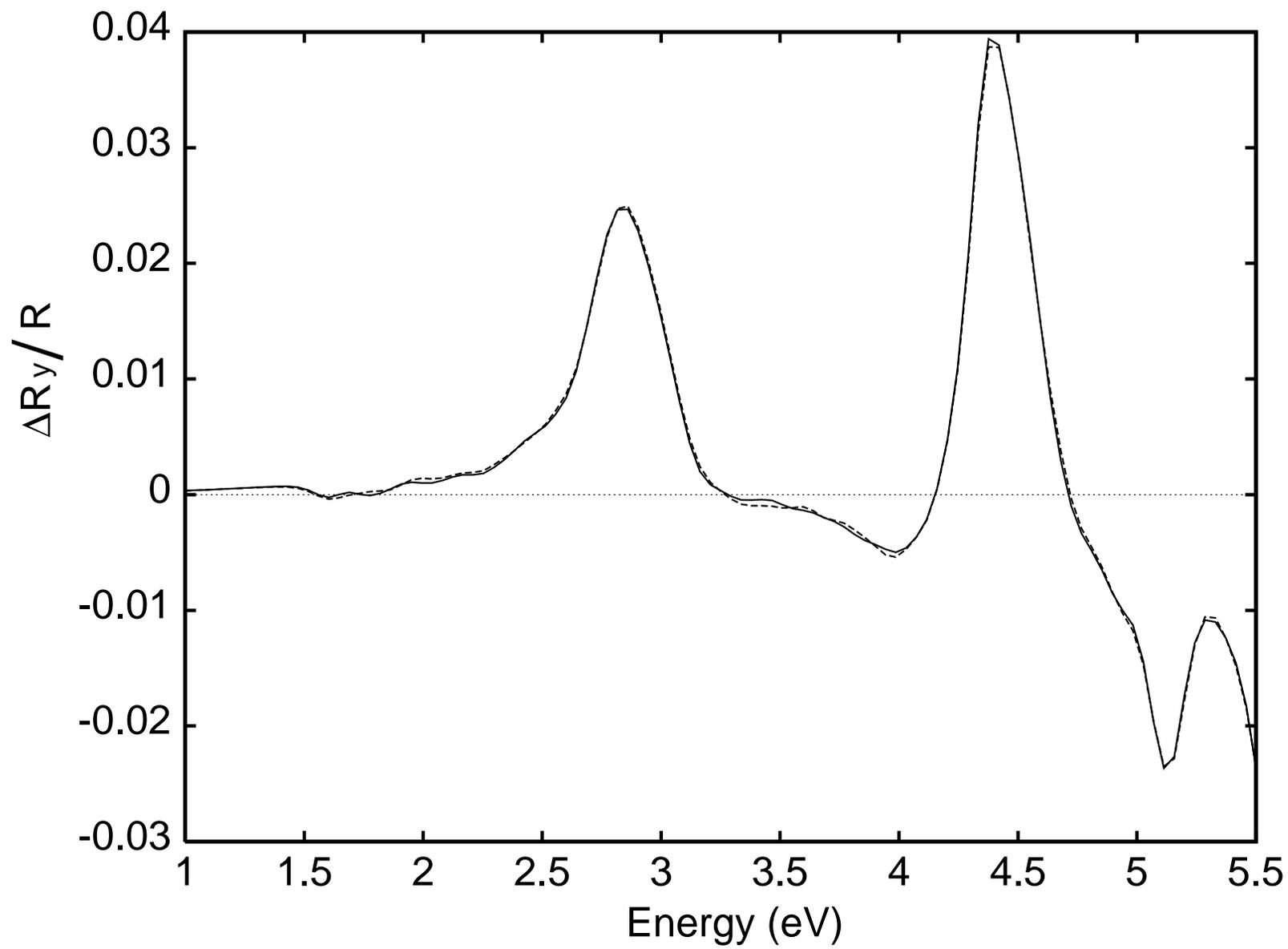
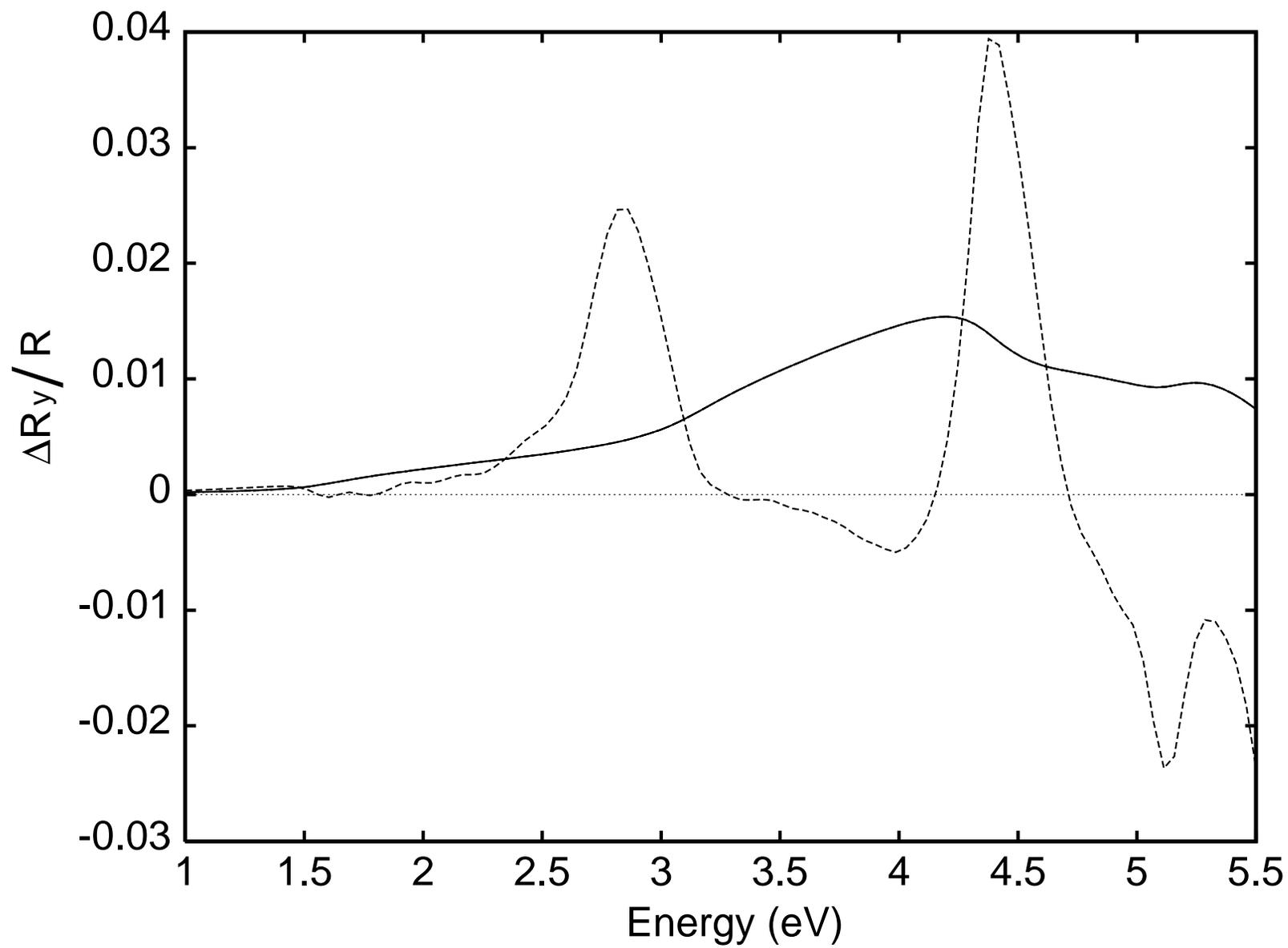
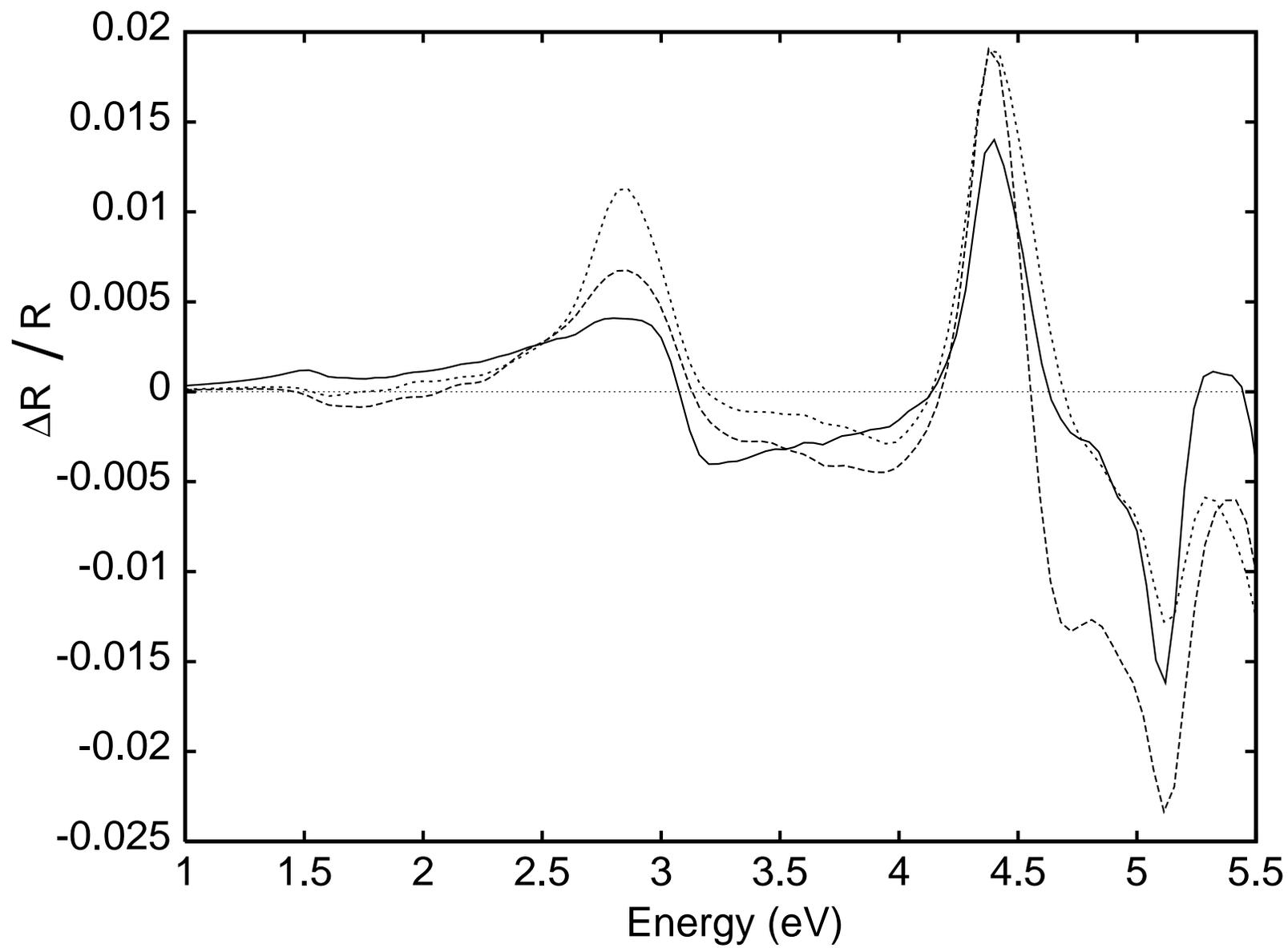
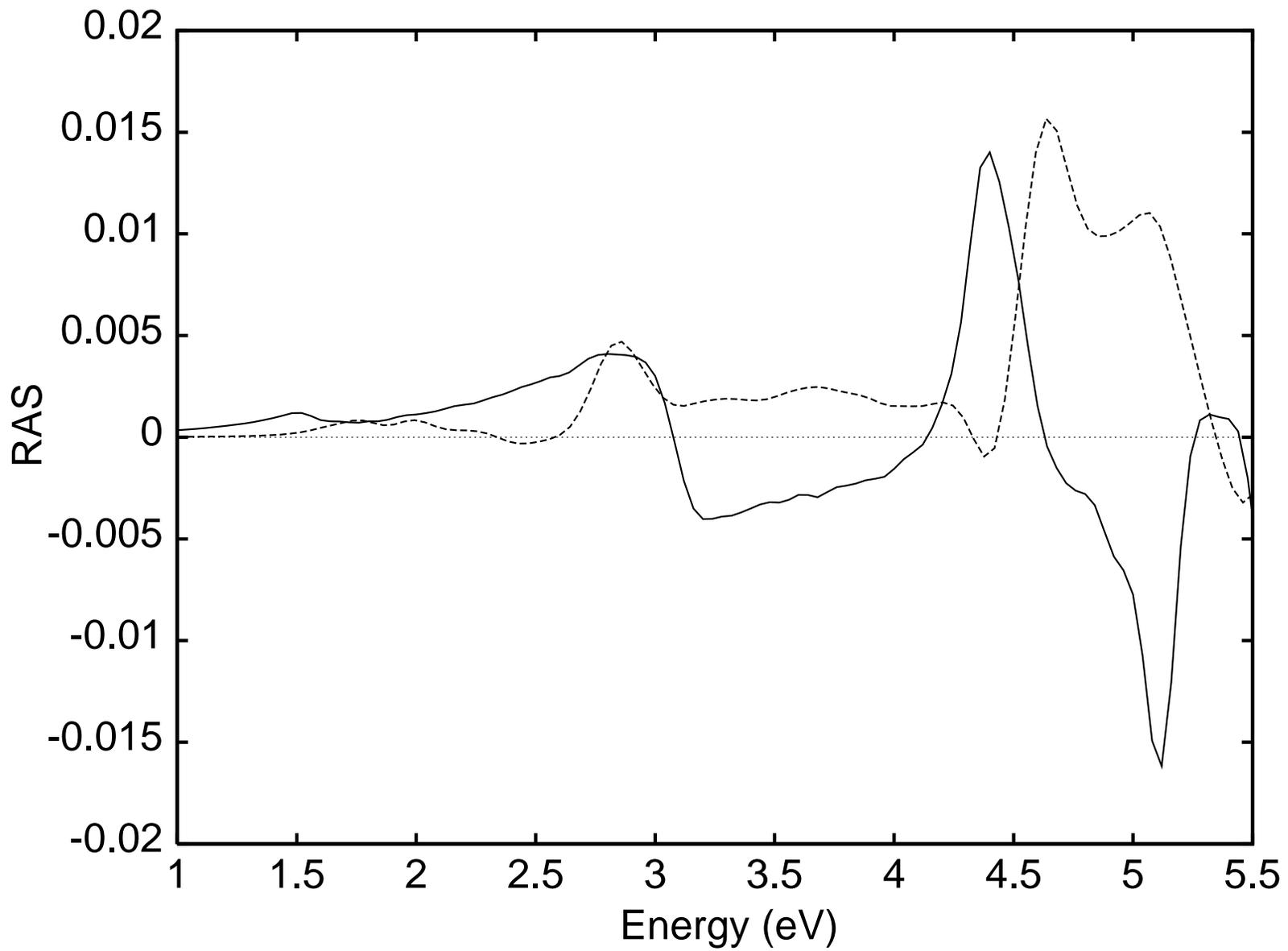
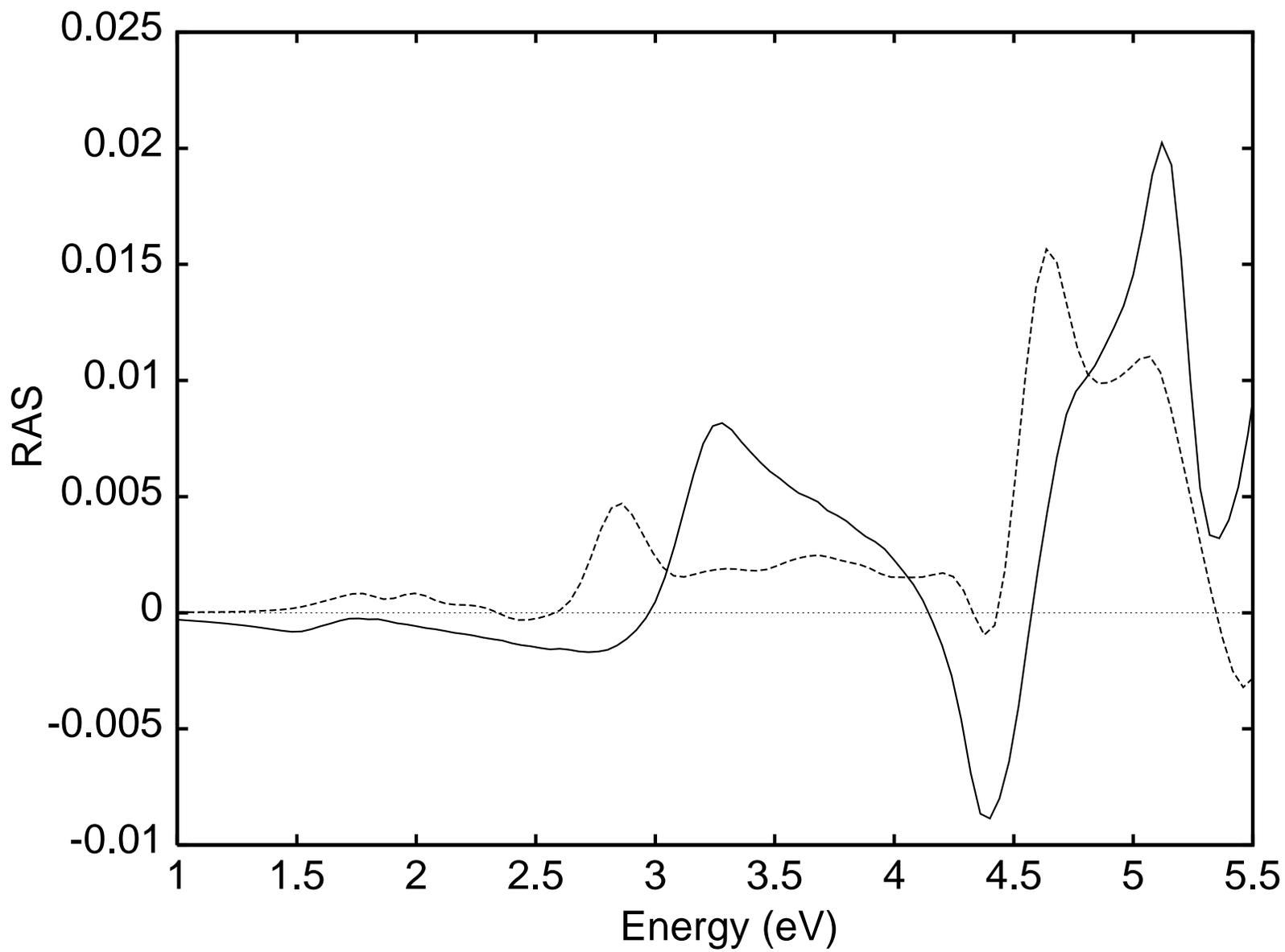
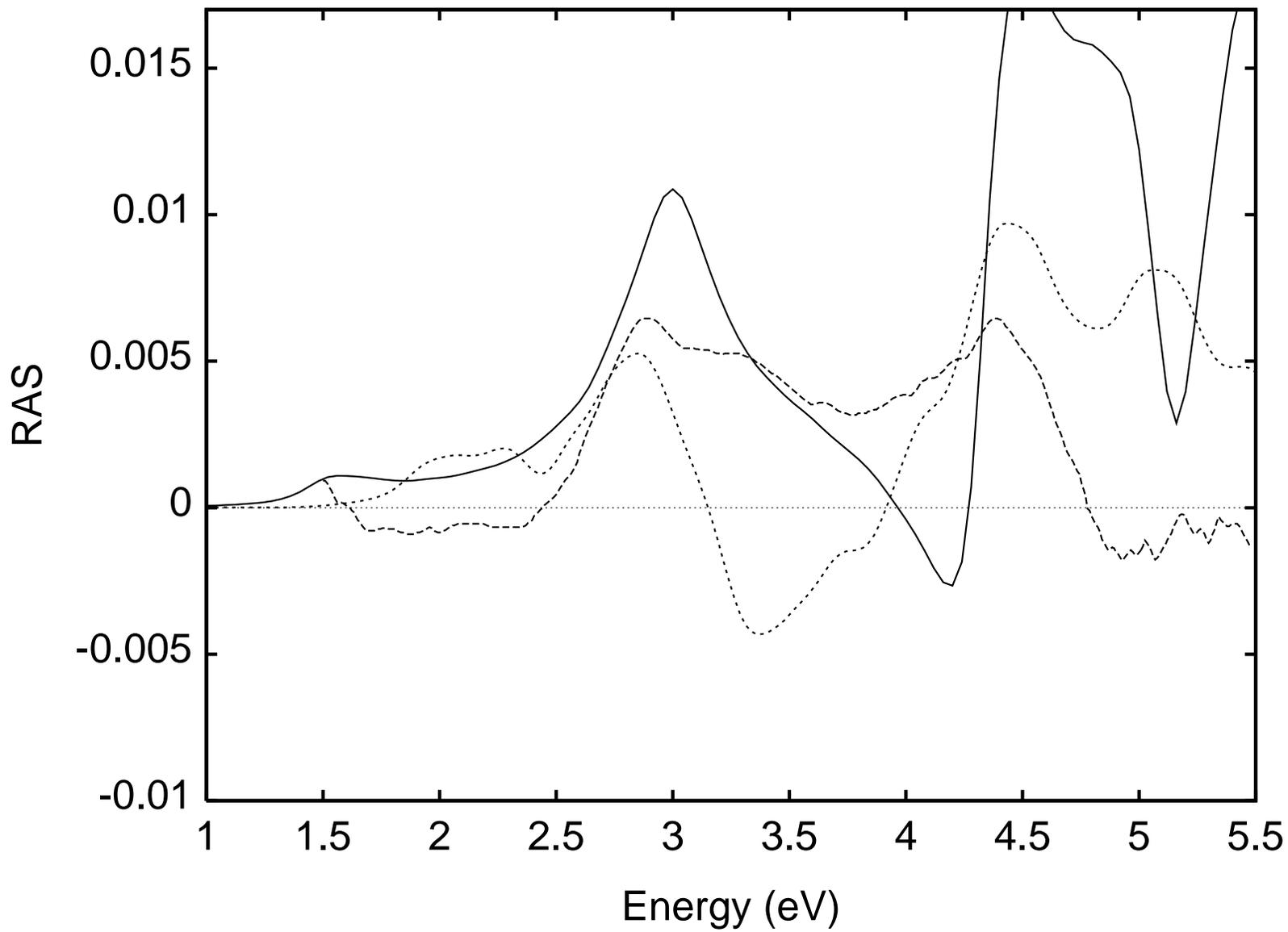
\end{document}